%% file: paper_9.tex
\documentclass[runningheads]{llncs}
\usepackage[T1]{fontenc}

\usepackage{graphicx,verbatim}
\usepackage{multicol, xcolor, colortbl}
\usepackage{hyperref}
\usepackage{pdfpages}
\usepackage{censor}
\usepackage{booktabs} 
\usepackage{tabularx}
\usepackage{array}
\usepackage{orcidlink}

\begin{document}
\title{Discriminating Distal Ischemic Stroke from Seizure-Induced Stroke Mimics Using Dynamic Susceptibility Contrast MRI}

%

\author{Marijn Borghouts\inst{1}~\orcidlink{0009-0002-3820-3957}, Richard McKinley\inst{2}~\orcidlink{0000-0001-8250-6117} \and Manuel Köstner\inst{2} \and Josien Pluim\inst{1} \and Roland Wiest\inst{2}~\orcidlink{0000-0001-7030-2045} \and Ruisheng Su\inst{1}~\orcidlink{0000-0002-5013-1370}}
\index{Borghouts, Marijn; McKinley, Richard; Köstner, Manuel; Pluim, Josien; Wiest, Roland; Su, Ruisheng}
\authorrunning{M. Borghouts et al.} 
\institute{
    $^1$Department of Biomedical Engineering, Eindhoven University of Technology, Eindhoven, The Netherlands \\
    $^2$Department of Diagnostic and Interventional Neuroradiology, Inselspital, University of Bern, Bern, Switzerland \\
    Correspondence: \email{m.m.borghouts@tue.nl}
}
\titlerunning{Discriminating stroke from stroke mimics using DSC MRI}

\maketitle
\begin{abstract}
\input{text/0.Abstract}


\end{abstract}

\section{Introduction}
\input{text/1.Introduction}

\section{Methodology}
\input{text/2.Methodology}

\section{Results}\label{sec:results}
\input{text/3.Results}

\section{Discussion and Conclusion}
\input{text/4.Discussion_and_Conclusion}

%
%
%
\bibliographystyle{splncs04} 
\bibliography{paper_9}

\newpage
\section{Appendices}
\input{text/5.Appendices}

\end{document}

%% file: text/0.Abstract.tex
Distinguishing acute ischemic strokes (AIS) from stroke mimics (SMs), particularly in cases involving medium and small vessel occlusions, remains a significant diagnostic challenge. While computed tomography (CT) based protocols are commonly used in emergency settings, their sensitivity for detecting distal occlusions is limited. This study explores the potential of magnetic resonance perfusion (MRP) imaging as a tool for differentiating distal AIS from epileptic seizures, a prevalent SM. Using a retrospective dataset of 162 patients (129 AIS, 33 seizures), we extracted region-wise perfusion map descriptors (PMDs) from dynamic susceptibility contrast (DSC) images. Statistical analyses identified several brain regions, located mainly in the temporal and occipital lobe, exhibiting significant group differences in certain PMDs. Hemispheric asymmetry analyses further highlighted these regions as discriminative. A logistic regression model trained on PMDs achieved an area under the receiver operating characteristic (AUROC) curve of 0.90, and an area under the precision recall curve (AUPRC) of 0.74, with a specificity of 92\% and a sensitivity of 73\%, suggesting strong performance in distinguishing distal AIS from seizures. These findings support further exploration of MRP-based PMDs as interpretable features for distinguishing true strokes from various mimics. The code is openly available at our GitHub \href{https://github.com/Marijn311/PMD_extraction_and_analysis}{github.com/Marijn311/PMD\_extraction\_and\_analysis}

%% file: text/1.Introduction.tex
Acute ischemic stroke (AIS) is a critical condition caused by a blocked blood vessel, depriving brain tissue of oxygen. Such occlusions can lead to irreversible brain damage, often resulting in permanent disability or death. In some patients, however, similar neurological symptoms arise from pathophysiologically distinct conditions. These conditions are known as stroke mimics (SMs) and commonly include peripheral vestibular disease, epileptic seizures, migraines, functional neurological disorders, and metabolic disturbances among others~\cite{Pohl2021}.

According to the 2025 update of the Heart Disease and Stroke Statistics report~\cite{Seth2025}, the global prevalence of ischemic stroke is nearly 70 million, with approximately 691,000 new and recurrent AIS cases each year in the United States alone. The direct annual medical cost of ischemic stroke and transient ischemic attack (TIA) in the U.S. is approximately \$25 billion~\cite{Seth2025}. Notably, up to as high as 50\% of suspected AIS cases are ultimately diagnosed as SMs~\cite{Buck2021,Hansson2019,Sequeira2016}, underscoring the importance of rapid and accurate diagnosis to avoid harmful or unnecessary interventions. 

Currently, non-contrast CT (NCCT) and CT angiography (CTA) are the most widely used imaging protocols for AIS diagnosis~\cite{Abdalkader2023}. However, CT imaging has limitations, particularly in detecting more distal occlusions~\cite{Stebner2025}. In a recent study~\cite{Duvekot2021}, medium vessel occlusion (MeVO) detection on CTA was compared between local observers at eight stroke centers and a centralized core laboratory. Local observers achieved a sensitivity of only 62\% compared to the reference core lab. In another study~\cite{Alotaibi2024}, they showed that nine starting practitioners achieved only 52\% accuracy in detecting MeVOs using CTA compared to a consensus between an experienced neuroradiologist and a stroke neurologist. 

MRP offers detailed insights into cerebral hemodynamics, enabling an evaluation of detailed perfusion patterns in the brain. This information could potentially improve the accuracy of distinguishing distal strokes from mimics. However, current literature offers rather limited information on MRP patterns specific to stroke mimics~\cite{Khalili2022}.

This study aims to address the existing gap by analyzing cerebral perfusion maps from patients with distal AIS and from patients with epileptic seizures, a common SM~\cite{Pohl2021}. Following the region-wise approach of~\cite{Kostner2023}, we extract statistical descriptors from perfusion map histograms. These perfusion map descriptors (PMDs) serve as explainable imaging features. Our goals are (1) to identify PMDs that distinguish distal AIS from mimics, and (2) to evaluate whether a machine learning (ML) model can classify these conditions based on PMD data.

%% file: text/2.Methodology.tex
\subsection{Data}
The dataset utilized in this study, consists of the diagnostic MR imaging that was performed on suspected stroke patients at the Inselspital in Bern, Switzerland. Following the Inselspital's protocols for suspected strokes, these patients received a DSC scan. Images were acquired on four Siemens MRI scanners (two 1.5T Aera/Avanto; two 3T Verio/Prisma) between 2008–2018. The DSC was acquired with a 2D EPI sequence for perfusion analysis. Images were made with a read FoV of 230 mm, a phase FoV of 100\%, a voxel size of 1.8 × 1.8 × 5.0 mm, a flip angle of 90°, and 80 repetitions, following an injection of 0.1 mmol/kg of gadolinium contrast agent with a flow rate of 5 mL/s. After the final diagnosis, the patients were weakly labeled with this diagnosis, i.e., stroke or seizure. 

We adopted the definition for distal (medium and small) vessel occlusions from the DISTAL trial~\cite{Marios2024}. Following these selection criteria, the final stroke dataset comprised 129 patients. The occlusion distribution over the vessel segments can be seen in Table~\ref{tab:distal_occlusion_distribution}. For the seizure patients an existing dataset of 33 cases was used. All the patient data originated from pre-curated datasets at the Inselspital and no additional exclusions were made in this study. The stroke cohort had a mean age of 68 years (standard deviation of 14), whereas the seizure cohort had a mean age of 50 years (standard deviation of 20). The proportion of male patients was 68\% in the stroke group and 58\% in the seizure group.


\begin{table}[ht]
\centering
\caption{Distribution of distal vessel occlusions in the final stroke dataset.}
\label{tab:distal_occlusion_distribution}
\begin{tabular}{llc}
\hline
\textbf{Artery} & \textbf{Segment(s)} & \textbf{Number of Patients} \\
\hline
Middle Cerebral Artery (MCA)   & M2                & 25 \\
                               & M3 or more distal & 56 \\
Anterior Cerebral Artery (ACA) & A1                & 0 \\
                               & A2 or more distal & 8 \\
Posterior Cerebral Artery (PCA)& P1                & 14 \\
                               & P2 or more distal & 26 \\
\hline
\textbf{Total}                 &                   & \textbf{129} \\
\hline
\end{tabular}
\end{table}

\subsection{Feature Extraction}
\begin{figure}[!t]
\includegraphics[width=\textwidth]{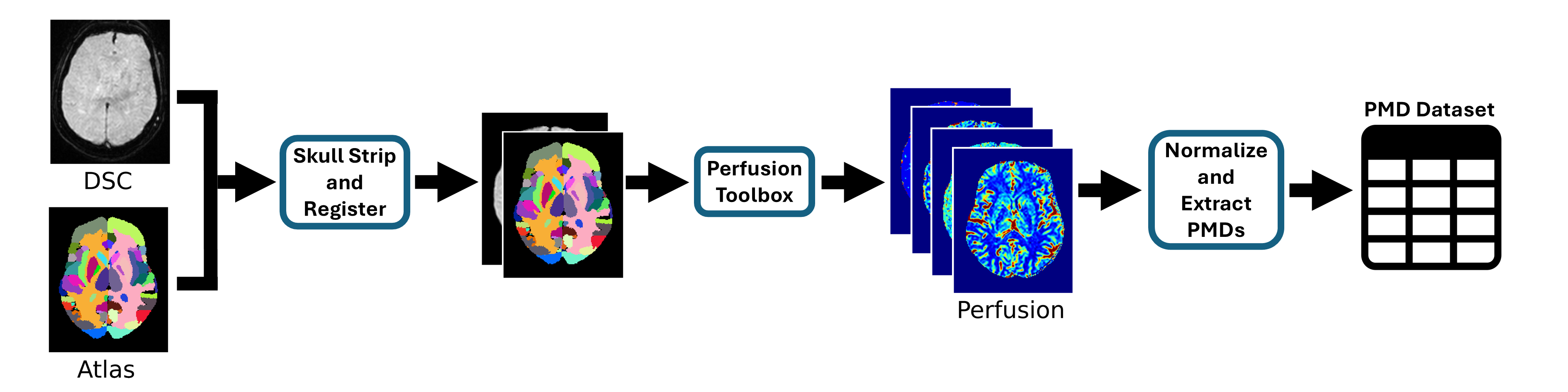}
\caption{Overview of the PMD extraction pipeline. The process begins with skull stripping to extract the brain from the MRP image. A brain atlas is then registered to the MRP image via its T1-weighted template. Perfusion maps are automatically generated from the DSC image using an open-source perfusion toolbox, with atlas guidance to ensure consistency across patients. The perfusion maps are normalized using the mean signal intensity, and finally, interpretable perfusion map descriptors (PMDs) are extracted as image features.}
\label{fig:graphical_abstract}
\end{figure}

To derive interpretable imaging features from the DSC scans, we developed a multi-step pipeline, as illustrated in Figure~\ref{fig:graphical_abstract}. Skull stripping was performed on the raw DSC volumes using the HD-BET tool~\cite{Isensee2019}. As this tool requires 3D input, the first time point of each 4D DSC sequence was used to generate the brain mask. This mask was then applied uniformly across all time points, as extracting each of the 80 volumes individually would drastically increase processing time. Next, a brain atlas was aligned to the DSC image by first registering its associated template to the DSC image. The resulting transformation matrix was then applied to warp the atlas accordingly. Registration was carried out using the Advanced Normalization Tools~\cite{Tustison2021} with the Symmetric Normalization algorithm~\cite{Avants2008} (which is based on a diffeomorphic and hence non-linear transformation). The brain atlas employed in this work was a composite of the Harvard-Oxford cortical and subcortical structural atlases~\cite{HO_atlas2006} as found in FSL~\cite{Jenkinson2012}, which together delineate 49 cortical and 7 subcortical regions per hemisphere, plus the brainstem, resulting in 113 unique regions.

Perfusion maps were generated from the DSC-MRI data using an open-source MATLAB toolbox originally developed by~\cite{Peruzzo2011}, which was further adapted in this work. The primary enhancement involved replacing the semi-automated arterial input function (AIF) selection, with a fully automated method. In the original implementation, the toolbox generated an AIF for each slice. Users were required to manually select the best slice. To eliminate the need for manual intervention, we implemented additional heuristic rules that automatically select the most appropriate slice based on predefined quality criteria. This fully automated pipeline improves scalability and reproducibility across large patient cohorts. To standardize AIF extraction in our adaption, we restricted the search area to a predefined anatomical region—the cingulate gyrus—chosen for its relatively large size, central location, and proximity to major cerebral arteries.

This improved toolbox produced volumetric maps of cerebral blood flow (CBF), cerebral blood volume (CBV), mean transit time (MTT), and time to maximum (Tmax) for each patient. Truncated singular value decomposition was used as the deconvolution algorithm, using 20\% of the maximum singular value as the truncation threshold. All perfusion maps were normalized using the mean signal intensity across the entire brain volume. This avoids needing (semi-)manual reference regions that require clinical validation. Given that our analyses rely on relative perfusion differences rather than absolute quantification, normalization to the whole-brain mean provided a consistent and practical solution. A comparison between the outputs of this open-source method and a commercial software package (Olea Sphere 3.0) is presented in Appendix~\ref{sec:foss_vs_commercial}.

We subsequently computed seven statistical measures (mean, median, standard deviation, interquartile range, skewness, kurtosis, and Hartigan’s dip) from the voxel intensity histograms within each of the 113 brain regions. This process was repeated for each of the four perfusion maps, resulting in a total of 7 × 113 × 4 = 3,164 PMDs. These PMDs served as the foundation for all subsequent analyses.

The proposed pipeline is computationally efficient, requiring no model training or user input. A dataset folder can be provided, and the entire dataset is processed automatically with a single command. The system is lightweight enough to run on a mobility laptop. Processing a single 256×256×19×80 image takes approximately 3 minutes on an Intel Core i5 CPU. Perfusion processing in MATLAB accounts for 25 seconds, while preprocessing steps—including reorientation, extraction of the first time point, perfusion map normalization, PMD extraction, and output saving—require an additional 30 seconds. Skull stripping using HD-BET on CPU takes 2 minutes but can be reduced to a few seconds with GPU acceleration. With access to a GPU and a moderately more powerful CPU, total processing time per patient can be reduced to under one minute.

\subsection{Experiments}
The first analysis assessed which PMDs differed significantly in distribution between patients with distal stroke and those with seizures. For each distribution comparison, the Wilcoxon rank-sum test was applied to measure significance. Bonferroni correction was used to adjust $p$-values for multiple comparisons across the 113 regions. Effect sizes were calculated using Cohen’s $d$. PMDs with an adjusted $p$-value below 0.05 and an absolute effect size above 0.3 were regarded as significant. A few examples of significantly different distributions are shown in Figure~\ref{fig:boxplot_example}.

The second analysis assessed hemispheric asymmetry for each PMD by calculating the absolute difference between each region and its contralateral counterpart. For example, the standard deviation of Tmax in the left lingual gyrus was compared to that in the right lingual gyrus. The distributions of these asymmetry measures were then compared between the distal stroke and seizure groups using the Wilcoxon rank-sum test and Cohen’s $d$ test. Due to bilateral pairing, only 56 unique region pairs were included in this analysis, with one region (brainstem) left out due to a lack of a counterlateral counterpart. Bonferroni correction was applied, correcting for 56 regions. PMDs with an adjusted $p$-value below 0.05 and an absolute effect size larger than 0.3 were regarded as significant. 

Lastly, a logistic regression classifier was trained to distinguish between distal stroke and seizure cases based on the tabulated PMD dataset. Logistic regression was chosen for its simplicity, explainability, and ability to handle tabular data. The model was implemented using the scikit-learn library with default hyperparameters, and the class weight parameter set to "balanced" to account for class imbalance. Alternative balancing techniques—including SMOTE, Gaussian noise augmentation, and minority oversampling—were evaluated. However, they did not yield performance improvements. Prior to training, all features were standardized using Z-score normalization, and missing values were imputed with feature-wise medians. Model evaluation was performed using leave-one-out cross-validation to make efficient use of the available dataset. Performance metrics included the receiver operating characteristic (ROC) curve, precision-recall (PR) curve, confusion matrix, and summary statistics. 95\% confidence intervals were calculated via 5000 bootstrap resamples. Exploratory analysis of alternative machine learning models can be found in Appendix~\ref{sec:other_ml_models}

%% file: text/3.Results.tex
\begin{figure}[ht]
\centering
\includegraphics[width=\textwidth]{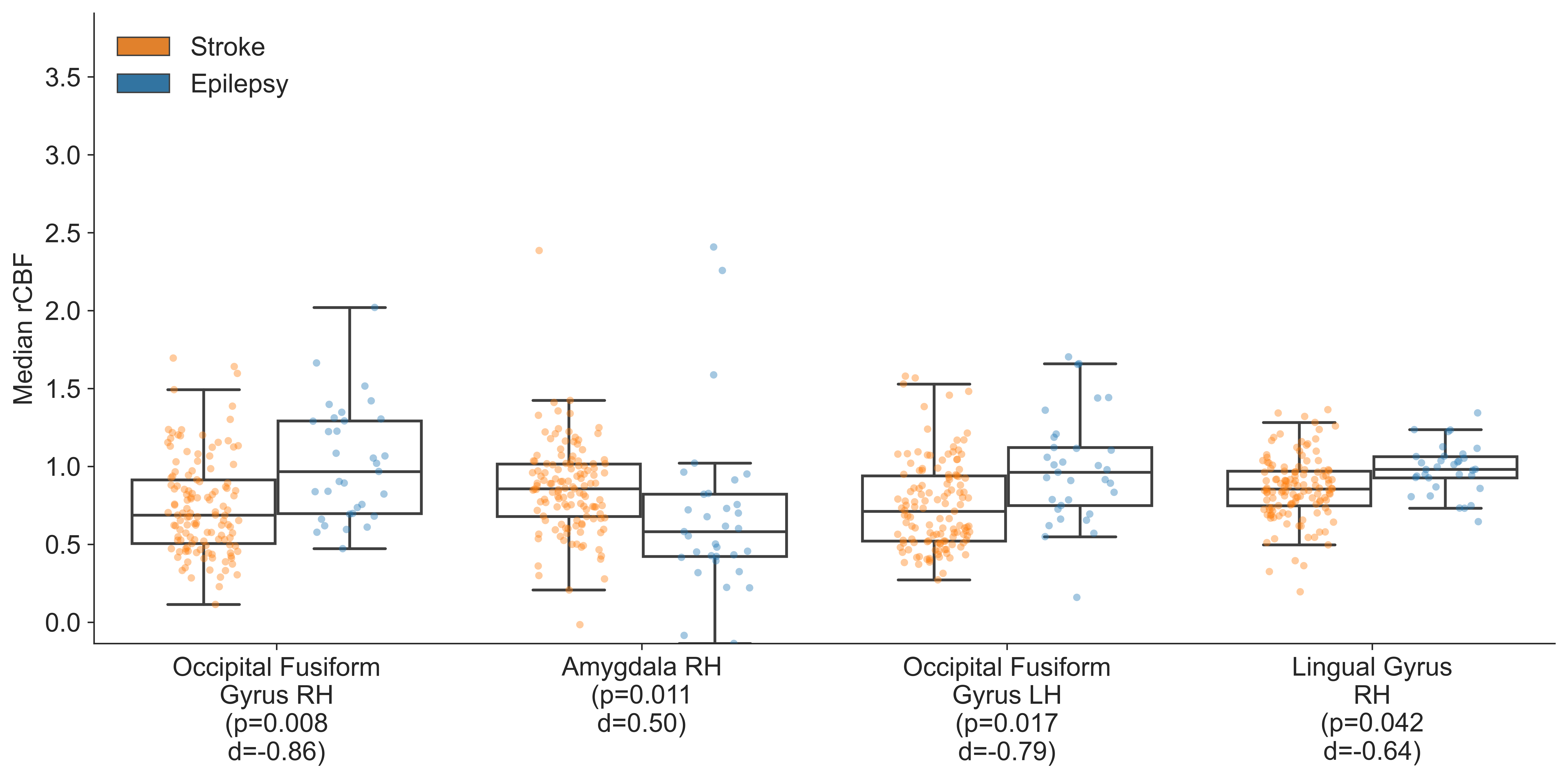}
\caption{Boxplots showing cerebral regions with significant (p < 0.05 and |d| > 0.3) group differences in median CBF. LH: left hemisphere; RH: right hemisphere}
\label{fig:boxplot_example}
\end{figure}

\begin{table}[ht]
\caption{Cerebral regions showing significant group difference (p < 0.05 and |d| > 0.3) in two or more PMDs. Regions are ranked top-to-bottom by the number of PMDs exhibiting significant differences.}
\label{tab:boxplot_summary}
\centering
{\fontsize{8pt}{8pt}\selectfont
    \begin{tabular}{|l|l|l|l|l|}
        \hline
        \textbf{REGION}& \textbf{CBF} & \textbf{CBV} & \textbf{MTT}  &\textbf{TMAX}\\
        \hline
        \textbf{Occipital Fusiform}& Mean, Median, SD, & Mean, Median, IQR& Skew&-\\
        \textbf{Gyrus}& IQR& & &\\
        \hline
        \textbf{Temporal Fusiform}& SD, IQR& SD, IQR& -&-\\
        \textbf{Cortex}& & & &\\
        \hline
        \textbf{Hippocampus}& -& Mean, Median& Mean, Median&-\\
        \hline
        \textbf{Lingual Gyrus}& Mean, Median& -& Skew&-\\
        \hline
        \textbf{Temporal Pole}& IQR, Skew& Kurtosis& -&-\\
        \hline
        \textbf{Parahippocampal}& IQR, Skew& Skew& -&-\\
        \textbf{Gyrus}& & & &\\
        \hline
        \textbf{Brainstem}& SD, IQR& -& -&-\\
        \hline
    \end{tabular}
}
\end{table}

Table \ref{tab:boxplot_summary} summarizes the brain regions that showed two or more significant PMD distribution differences between the distal AIS and seizure groups. The table ranks cerebral regions by the number of significant PMDs associated with each. The statistical metrics, such as the median or kurtosis, mostly served to quantify images for compatibility with a computer-based algorithm. We understand that clinicians are not accustomed to analyzing features such as the IQR in a specific brain region. However, the underlying rationale for this organization is that a greater number of differing PMDs within a given region suggests more pronounced visual differences in that region. Hence, our results suggest which brain regions are likely to be visually distinct between patient groups and are thus of interest to clinicians. Notably, among the 27 PMDs highlighted in Table~\ref{tab:boxplot_summary}, only seven unique cerebral regions were identified. This observation suggests that specific regions consistently emerge as relevant across multiple PMDs, underscoring their potential importance in distinguishing between stroke and seizure.

Among the identified regions, the occipital fusiform gyrus emerged as the most discriminative for differentiating AIS from seizures, with eight PMDs exhibiting significant differences in this region. Other cerebral regions associated with multiple significant PMDs included the temporal fusiform cortex and the hippocampus (each with four PMDs), as well as the lingual gyrus, temporal pole, and parahippocampal gyrus (each with three PMDs), and lastly the brainstem with two PMDs. Notably, the CBF and CBV maps yielded a greater number of PMDs with significant differences (fourteen and nine, respectively) compared to the MTT and Tmax maps (four and zero PMDs). An extended overview of all significant PMDs, including lateralization across hemispheres and the corresponding $p$- and $d$-values, is provided in Appendix~\ref{sec:extended_results}.

\begin{table}[ht]
\caption{Cerebral regions showing significant group difference (p < 0.05 and |d| > 0.3) in the asymmetry of one or more PMDs.}
\label{tab:asymmetry_results}
\centering
{\fontsize{8pt}{8pt}\selectfont
    \begin{tabular}{|l|l|l|l|l|}
        \hline
        \textbf{REGION}& \textbf{CBF} & \textbf{CBV} & \textbf{MTT}  &\textbf{TMAX}\\
        \hline
        \textbf{Temporal Fusiform Cortex}& Mean, Median, IQR& SD, IQR& -&-\\
        \hline
        \textbf{Precuneous Cortex}& Median, SD, IQR& -& -&-\\
        \hline
        \textbf{Middle Temporal Gyrus}& Median& Median& IQR&-\\
        \hline
        \textbf{Temporal Pole}& Median& Median& -&-\\
        \hline
        \textbf{Frontal Pole}& Median& IQR& -&-\\
        \hline
    \end{tabular}
}
\end{table}

As shown in Table \ref{tab:asymmetry_results}, several brain regions demonstrated two or more statistically significant differences in PMD asymmetry between the distal AIS and seizure groups. These include the temporal fusiform cortex, precuneus cortex, middle temporal gyrus, temporal pole, and frontal pole.

\begin{figure}
\includegraphics[width=\textwidth]{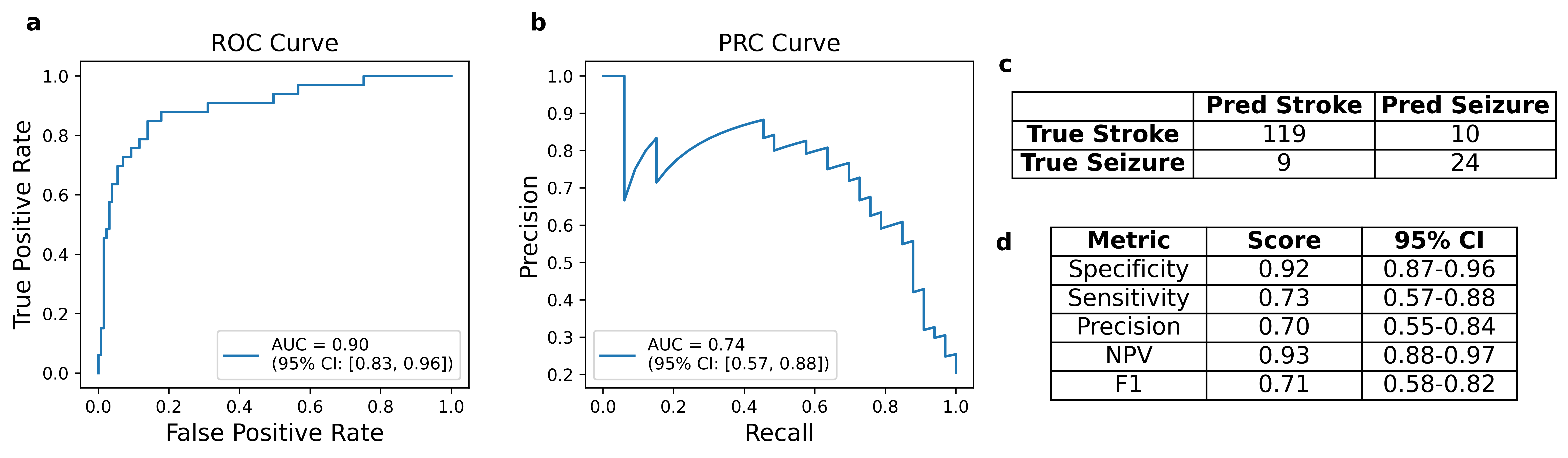}
\caption{Performance metrics of the logistic regression classifier obtained using leave-one-out cross-validation.}
\label{fig:ml_metrics}
\end{figure}

Figure \ref{fig:ml_metrics} summarizes the performance of the logistic regression model. The receiver operating characteristic (ROC) curve (subplot a) demonstrates strong overall discriminative performance, with an area under the curve (AUC) of 0.90 (95\% CI: 0.83-0.96). It should be noted though that the ROC curve is not the most sensitive to misclassified minority cases in imbalanced datasets. The precision-recall (PR) curve (subplot b) yielded a lower AUC of 0.74 (95\% CI: 0.57-0.88), better reflecting the impact of class imbalance and the model’s comparatively reduced ability to detect seizure cases. This phenomenon is also illustrated in the confusion matrix (subplot c), which shows that the model correctly identified 119 out of 129 true stroke cases, corresponding to a specificity of 92\%. Among the 33 true seizure cases, 24 were correctly classified, giving a sensitivity of 73\%. Additional classification metrics with confidence intervals are provided in subplot d. A SHap-ley Additive exPlanations analysis can be found in Appendix~\ref{sec:shap}.

%% file: text/4.Discussion_and_Conclusion.tex
This study aimed to investigate whether perfusion MRI data could be used to distinguish medium- to small-vessel strokes from seizures, a common stroke mimic. By extracting PMDs from volumetric perfusion images, we derived a set of interpretable image features. Statistical analyses were conducted to identify cerebral regions where PMD distributions differed significantly between stroke and seizure groups. In addition, a logistic regression model was trained to assess the discriminatory power of the extracted PMDs.

Key findings from our analyses indicate that specific brain regions, primarily within the temporal and occipital lobes, exhibit significant differences in cerebral perfusion patterns between patients with distal AIS and those with seizures. Hemispheric asymmetry analysis further reinforced the discriminative value of these two lobes. This insight may aid clinicians in identifying relevant regions of interest when diagnostic uncertainty exists between distal stroke and seizure. 

It is well established that up to two-thirds of partial seizures originate in the temporal lobe \cite{Semah1998}, making our findings regarding the temporal lobe and adjacent occipital lobe unsurprising. While these results may not be novel from a pathophysiological perspective, it is noteworthy that these patterns can be detected using MRP, a modality that has barely been explored for SMs. In contrast, distal strokes can impact a wide range of brain regions depending on the location of the vascular occlusion. This heterogeneity makes it more challenging to identify consistent regional patterns or specific perfusion map descriptors (PMDs) characteristic of distal strokes. However, more advanced machine learning approaches may be capable of uncovering complex, non-linear relationships among PMDs that are indicative of distal stroke. We believe this represents a promising direction for future research.

The logistic regression model trained on PMD features achieved an area under the precision recall curve of 0.74, demonstrating strong discriminative performance in distinguishing medium- to small-vessel AIS from seizure cases in imbalanced datasets. These results underscore the potential of the PMDs extraction, facilitating robust statistical analysis and machine learning applications—even in studies with limited dataset sizes. The model demonstrated high specificity (92\%), suggesting its potential as a clinical decision‑support tool, especially useful for ruling out stroke mimics in equivocal cases. However, the primary aim of this model was not to develop a deployable classifier, but rather to evaluate whether the PMD features extracted from perfusion MRI carry sufficient discriminative information. 

Despite limitations, including a relatively small dataset, single-center data, and retrospective design, the findings provide encouraging preliminary evidence. They support further research involving the application of more advanced machine learning techniques, expansion to larger and more diverse datasets, inclusion of additional mimic subtypes, and integration of clinical and demographic variables. Future studies might also benefit from incorporating complementary imaging modalities such as diffusion-weighted imaging (DWI), exploring alternative tasks like lesion segmentation or outcome prediction, and investigating the use of CT perfusion for broader applicability. 

In conclusion, we developed a fully automated, interpretable, and openly available pipeline for extracting explainable image features from raw DSC-MRI data. The pipeline can be run without GPU acceleration, promoting accessibility and reproducibility. We demonstrated that these extracted image features can effectively differentiate between distal AIS and seizure cases. This provides evidence that MR perfusion imaging contains diagnostic information that may not be readily captured by standard NCCT/CTA. Moreover, our results suggest that the PMD pipeline captures a substantial portion of this relevant information to enable meaningful analyses. While this research remains preliminary and is not intended to inform immediate clinical practice, it highlights a promising direction. With further development and validation, this approach has the potential to contribute to the understanding of perfusion characteristics in stroke and its mimics, and ultimately improve diagnosis and clinical decision-making in stroke care.
\newline
\newline
\textbf{Acknowledgment.} This work is supported by the MIMIC project, funded by the NWO NGF AiNed XS Europe grant (NGF.1609.242.047).
\newline
\newline
\textbf{Disclosure of Interests.} The authors have no competing interests to declare that are relevant to the content of this article.

%% file: text/5.Appendices.tex

\subsection{Open-Source vs Commercial DSC Processing}\label{sec:foss_vs_commercial}
\begin{figure}[!htbp]
\includegraphics[width=\textwidth]{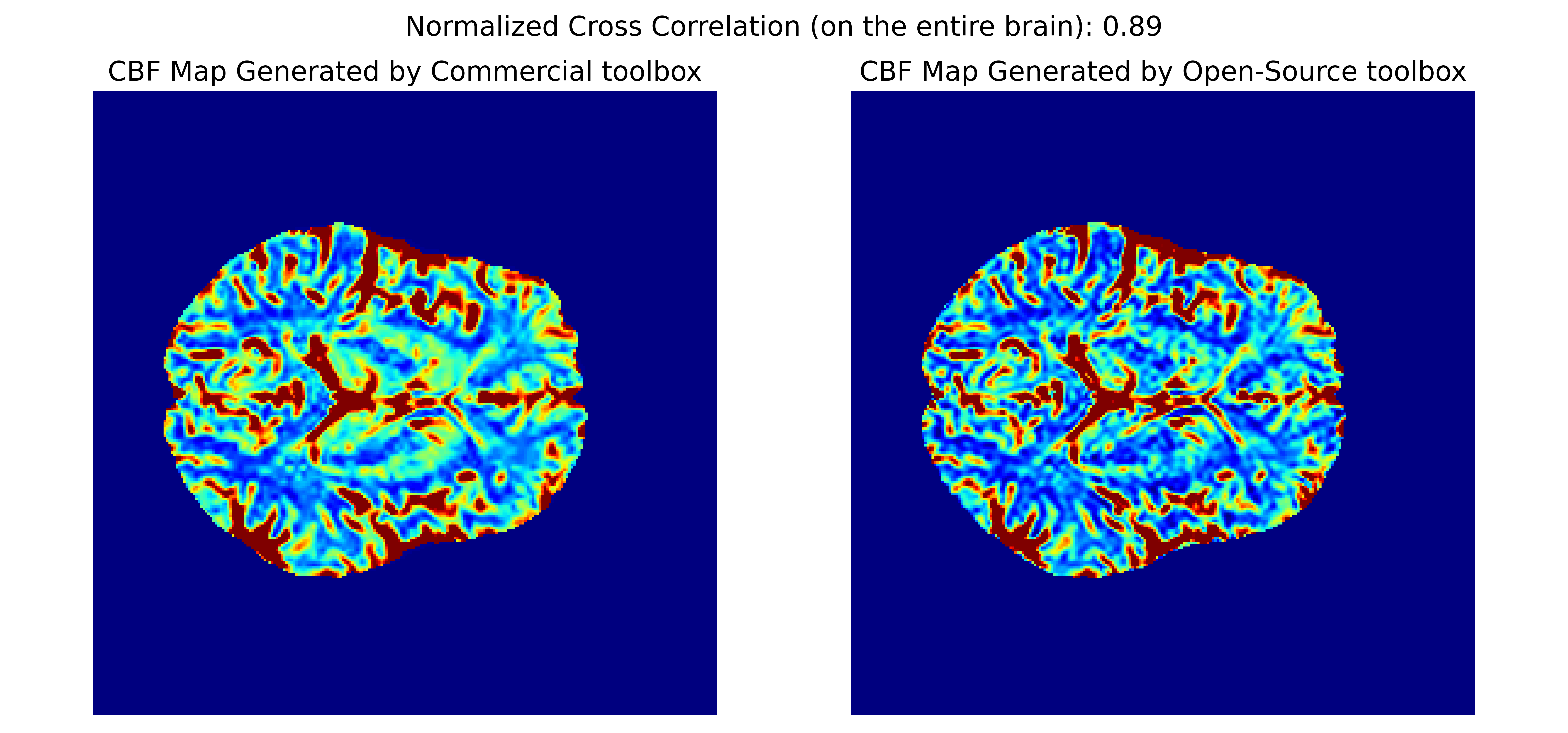}
\caption{Comparison of CBF maps generated by the commercial software Olea Sphere 3.0 and the open-source toolbox evaluated in this study, using data from Patient 1. The color scale was standardized using the 95\% intensity range to define the minimum and maximum values for the colorbar.}
\label{fig:foss_validation_cbf}
\end{figure}

\begin{figure}[!htbp]
\includegraphics[width=\textwidth]{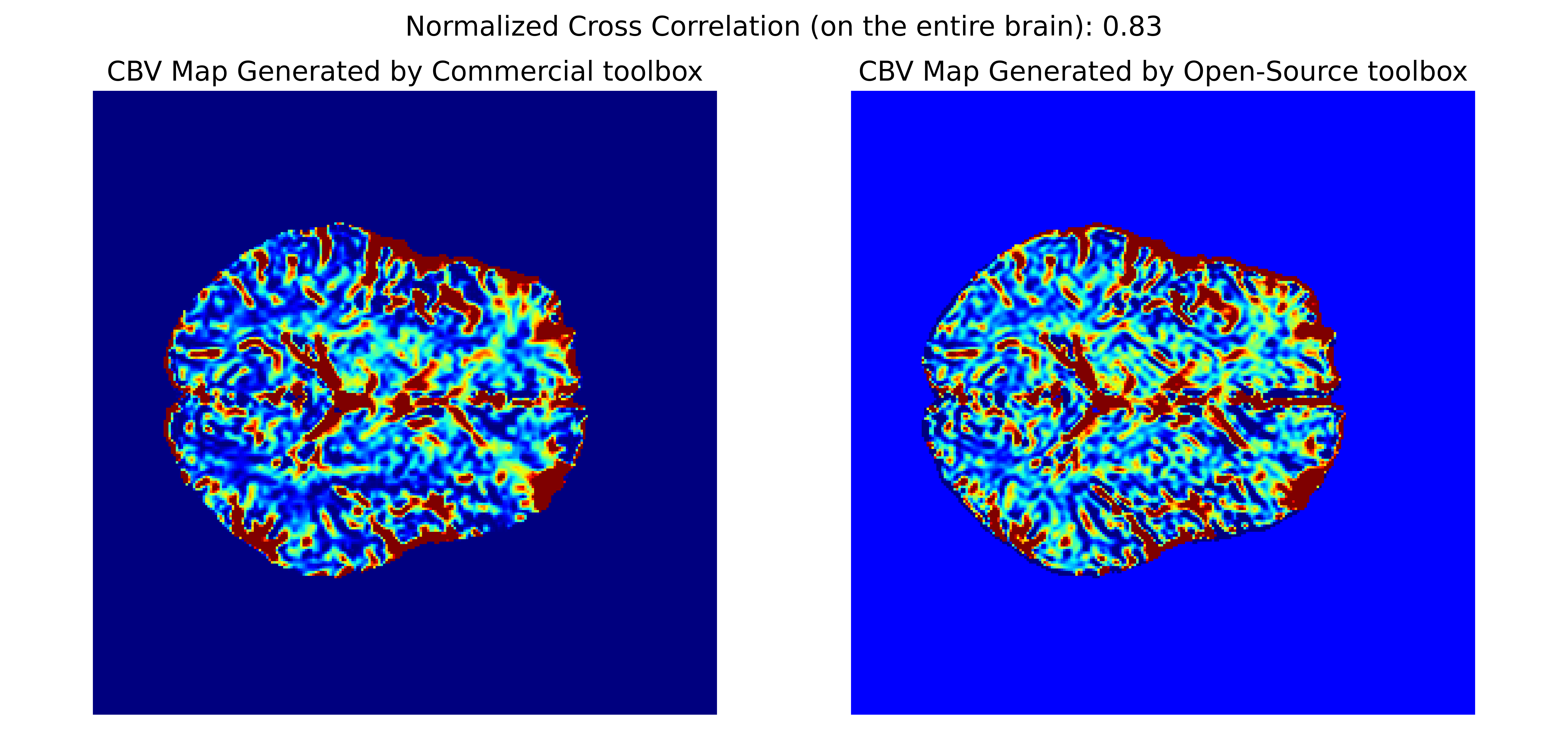}
\caption{Comparison of CBV maps generated by the commercial software Olea Sphere 3.0 and the open-source toolbox evaluated in this study, using data from Patient 1. The color scale was standardized using the 95\% intensity range to define the minimum and maximum values for the colorbar.}
\label{fig:foss_validation_cbv}
\end{figure}

Several commercial FDA-approved software packages are available for generating perfusion maps from DSC-MRI data. However, studies have shown that these tools can produce markedly different results~\cite{Korfiatis2016,Kudo2017}. These inconsistencies between commercial solutions complicate the validation of open-source alternatives, as there is no universally accepted gold standard. Nevertheless, side-by-side visual comparisons and quantitative similarity metrics can still offer insights. To evaluate the performance of the open-source toolbox used in this work, scans from eight stroke patients were analyzed. For each case, CBF and CBV maps generated by the commercial software Olea Sphere 3.0 were available. The same raw patient DSC data were processed using the open-source tool as it was used in the main work. The resulting perfusion maps were compared to those obtained from Olea Sphere. Figures~\ref{fig:foss_validation_cbf} and~\ref{fig:foss_validation_cbv} present a visual comparison between the two toolboxes, while Table~\ref{tab:foss_validation} reports the normalized cross-correlation values for all comparisons.

\begin{table}[!htbp]
\caption{Normalized cross-correlation (NCC) values comparing perfusion maps (CBF and CBV) generated by the commercial software Olea Sphere 3.0 and the open-source toolbox evaluated in this study.}
\label{tab:foss_validation}
\centering
{\fontsize{9pt}{10pt}\selectfont
\begin{tabular}{lcc}
\toprule
\textbf{Patient} & \textbf{CBF} & \textbf{CBV} \\
\midrule
Patient 1 & 0.89 & 0.83 \\
Patient 2 & 0.89 & 0.90 \\
Patient 3 & 0.73 & 0.64 \\
Patient 4 & 0.84 & 0.88 \\
Patient 5 & 0.95 & 0.92 \\
Patient 6 & 0.92 & 0.93 \\
Patient 7 & 0.88 & 0.92 \\
Patient 8 & 0.69 & 0.65 \\
\midrule
\textbf{Mean} & \textbf{0.85} & \textbf{0.83} \\
\textbf{Standard Deviation} & \textbf{0.09} & \textbf{0.11} \\
\bottomrule
\end{tabular}
}
\end{table}

\newpage
\subsection{Extended Results}
\label{sec:extended_results}
The following two full-page tables present the extended results from the distributional analyses of the PMDs. The first table expands upon Table \ref{tab:boxplot_summary} from the main paper's Results section and lists all PMDs that show statistically significant distributional differences between the stroke and seizure groups. The second table extends Table \ref{tab:asymmetry_results} and includes PMDs that exhibit significant hemispheric asymmetry differences between the two groups. In contrast to the main text, these extended tables provide hemisphere-specific results by distinguishing between the left hemisphere (LH) and right hemisphere (RH). For each reported PMD, we include both the $p$-value from the Wilcoxon rank-sum test and the corresponding effect size (Cohen’s $d$). All $p$-values were corrected for multiple comparisons using the Bonferroni method.



|  | CBF | CBV_LC | MTT | Tmax |
|---|---|---|---|---|
| Mean | Occipital Fusiform Gyrus LH (p:0.033, d:-0.697) |  |  |  |
|  | Lingual Gyrus RH (p:0.014, d:-0.899) | Occipital Fusiform Gyrus RH (p:0.014, d:-0.828) |  |  |
|  | Occipital Fusiform Gyrus RH (p:0.002, d:-1.002) | Hippocampus RH (p:0.001, d:0.942) | Hippocampus LH (p:0.005, d:0.619) |  |
| Median | Occipital Fusiform Gyrus LH (p:0.017, d:-0.790) |  |  |  |
|  | Lingual Gyrus RH (p:0.042, d:-0.640) |  |  |  |
|  | Occipital Fusiform Gyrus RH (p:0.008, d:-0.865) | Occipital Fusiform Gyrus RH (p:0.022, d:-0.871) |  |  |
|  | Amygdala RH (p:0.011, d:0.497) | Hippocampus RH (p:0.005, d:0.753) | Hippocampus LH (p:0.004, d:0.756) |  |
| STD | Temporal Fusiform Cortex, anterior division LH (p:0.027, d:-0.583) |  |  |  |
|  | Temporal Fusiform Cortex, anterior division RH (p:0.000, d:-0.673) |  |  |  |
|  | Temporal Fusiform Cortex, posterior division RH (p:0.001, d:-0.675) |  |  |  |
|  | Occipital Fusiform Gyrus RH (p:0.018, d:-0.929) | Temporal Fusiform Cortex, anterior division RH (p:0.002, d:-0.575) |  |  |
|  | Brain-Stem (p:0.008, d:-0.628) | Temporal Fusiform Cortex, posterior division RH (p:0.010, d:-0.314) |  |  |
| IQR | Temporal Pole LH (p:0.016, d:-0.836) |  |  |  |
|  | Temporal Fusiform Cortex, posterior division LH (p:0.008, d:-0.712) |  |  |  |
|  | Occipital Fusiform Gyrus LH (p:0.016, d:-0.712) |  |  |  |
|  | Parahippocampal Gyrus, anterior division RH (p:0.043, d:-0.569) |  |  |  |
|  | Temporal Fusiform Cortex, anterior division RH (p:0.003, d:-0.672) |  |  |  |
|  | Temporal Fusiform Cortex, posterior division RH (p:0.002, d:-0.663) |  |  |  |
|  | Occipital Fusiform Gyrus RH (p:0.000, d:-1.047) | Temporal Fusiform Cortex, anterior division RH (p:0.008, d:-0.594) |  |  |
|  | Parietal Operculum Cortex RH (p:0.022, d:-0.730) | Temporal Fusiform Cortex, posterior division RH (p:0.006, d:-0.399) |  |  |
|  | Brain-Stem (p:0.000, d:-0.679) | Occipital Fusiform Gyrus RH (p:0.006, d:-0.825) |  |  |
| Skewness | Temporal Pole RH (p:0.040, d:0.753) |  | Lingual Gyrus RH (p:0.019, d:-0.752) |  |
|  | Subcallosal Cortex RH (p:0.032, d:0.816) | Parahippocampal Gyrus, anterior division LH (p:0.003, d:0.791) | Occipital Fusiform Gyrus RH (p:0.031, d:-0.794) |  |
|  | Parahippocampal Gyrus, anterior division RH (p:0.034, d:0.867) | Parahippocampal Gyrus, anterior division RH (p:0.000, d:1.006) | Caudate LH (p:0.020, d:-0.737) |  |
| Kurtosis |  | Temporal Pole LH (p:0.040, d:0.735) |  |  |
| Unimodality |  |  |  |  |

|  | CBF | CBV_LC | MTT | Tmax |
|---|---|---|---|---|
| **Mean** | Temporal Fusiform Cortex, anterior division LH (p:0.000, d:-0.922) |  |  |  |
|  | Temporal Fusiform Cortex, anterior division RH (p:0.000, d:-0.922) |  |  |  |
| **Median** | Frontal Pole LH (p:0.018, d:-0.765) |  |  |  |
|  | Temporal Pole LH (p:0.001, d:-0.826) |  |  |  |
|  | Middle Temporal Gyrus, anterior division LH (p:0.010, d:-0.829) |  |  |  |
|  | Lateral Occipital Cortex, inferior division LH (p:0.026, d:-0.895) |  |  |  |
|  | Precuneous Cortex LH (p:0.034, d:-0.956) |  |  |  |
|  | Parahippocampal Gyrus, anterior division LH (p:0.010, d:-0.718) |  |  |  |
|  | Temporal Fusiform Cortex, anterior division LH (p:0.006, d:-0.918) |  |  |  |
|  | Frontal Pole RH (p:0.018, d:-0.765) |  |  |  |
|  | Temporal Pole RH (p:0.001, d:-0.826) |  |  |  |
|  | Middle Temporal Gyrus, anterior division RH (p:0.010, d:-0.829) |  |  |  |
|  | Lateral Occipital Cortex, inferior division RH (p:0.026, d:-0.895) | Temporal Pole LH (p:0.001, d:-0.758) |  |  |
|  | Precuneous Cortex RH (p:0.034, d:-0.956) | Middle Temporal Gyrus, anterior division LH (p:0.042, d:-0.352) |  |  |
|  | Parahippocampal Gyrus, anterior division RH (p:0.010, d:-0.718) | Temporal Pole RH (p:0.001, d:-0.758) |  |  |
|  | Temporal Fusiform Cortex, anterior division RH (p:0.006, d:-0.918) | Middle Temporal Gyrus, anterior division RH (p:0.042, d:-0.352) |  |  |
| **STD** | Superior Temporal Gyrus, posterior division LH (p:0.023, d:-0.837) |  |  |  |
|  | Inferior Temporal Gyrus, anterior division LH (p:0.028, d:-0.677) |  |  |  |
|  | Precuneous Cortex LH (p:0.044, d:-0.880) |  |  |  |
|  | Superior Temporal Gyrus, posterior division RH (p:0.023, d:-0.837) |  |  |  |
|  | Inferior Temporal Gyrus, anterior division RH (p:0.028, d:-0.677) | Temporal Fusiform Cortex, anterior division LH (p:0.026, d:-0.695) |  |  |
|  | Precuneous Cortex RH (p:0.044, d:-0.880) | Temporal Fusiform Cortex, anterior division RH (p:0.026, d:-0.695) |  |  |
| **IQR** | Precuneous Cortex LH (p:0.024, d:-0.987) |  |  |  |
|  | Temporal Fusiform Cortex, anterior division LH (p:0.000, d:-0.876) |  |  |  |
|  | Precuneous Cortex RH (p:0.024, d:-0.987) | Frontal Pole LH (p:0.026, d:-0.523) |  |  |
|  | Temporal Fusiform Cortex, anterior division RH (p:0.000, d:-0.876) | Temporal Fusiform Cortex, anterior division LH (p:0.000, d:-0.826) |  |  |
|  | Cerebral White Matter LH (p:0.019, d:-0.838) | Frontal Pole RH (p:0.026, d:-0.523) | Middle Temporal Gyrus, anterior division LH (p:0.050, d:-0.459) |  |
|  | Cerebral White Matter RH (p:0.019, d:-0.838) | Temporal Fusiform Cortex, anterior division RH (p:0.000, d:-0.826) | Middle Temporal Gyrus, anterior division RH (p:0.050, d:-0.459) |  |
| **Skewness** |  |  | Central Opercular Cortex LH (p:0.027, d:0.671) |  |
|  |  |  | Central Opercular Cortex RH (p:0.027, d:0.671) |  |
| **Kurtosis** |  |  |  |  |
| **Unimodality** |  |  |  |  |



\subsection{SHAP Value Analysis}
\label{sec:shap}
The logistic regression model was trained on thousands of PMDs, each representing a unique combination of perfusion image type, statistical descriptor, and brain region. To assess feature importance, we conducted a SHAP (SHapley Additive exPlanations) analysis~\cite{Lundberg2017}, computing a SHAP value for each PMD to quantify its contribution to the model's output.

PMDs were ranked by SHAP value, with the least influential assigned rank 1 and the most influential ranked highest. To evaluate the relative importance of each image type, statistical metric, and brain region, we averaged the ranks of all PMDs sharing a common component (e.g., all containing CBF). This process was repeated for each component category, and results are shown in Table~\ref{tab:avg_pmd_ranks}  (limited to the top 10 brain regions for brevity).

\begin{table}[ht]
\caption{Average SHAP-based rank of PMDs grouped by image type, statistical metric, and brain region. Higher ranks indicate higher importance. Max rank = 4580, Min rank = 1. Rounded to nearest integer.}
\label{tab:avg_pmd_ranks}
\centering
\small
\vspace{0.5em}

\begin{tabular}{|l|c|}
\hline
\textbf{Image Type} & \textbf{Average Rank} \\
\hline
CBF & 2829 \\
Tmax & 2560 \\
MTT & 2103 \\
CBV & 1670 \\
\hline
\end{tabular}

\vspace{1em}

\begin{tabular}{|l|c|}
\hline
\textbf{Statistical Metric} & \textbf{Average Rank} \\
\hline
Unimodality Asymmetry & 3070 \\
Unimodality & 2630 \\
Median & 2524 \\
Mean & 2379 \\
Median Asymmetry & 2353 \\
SD Asymmetry & 2280 \\
IQR Asymmetry & 2260 \\
Mean Asymmetry & 2221 \\
IQR & 1753 \\
SD & 1440 \\
\hline
\end{tabular}

\vspace{1em}

\begin{tabular}{|l|c|}
\hline
\textbf{Brain Region} & \textbf{Average Rank} \\
\hline
Brain-Stem & 2995 \\
Right Lateral Ventricle & 2936 \\
Right Central Operculum Cortex & 2929 \\
Right Thalamus & 2882 \\
Left Lateral Ventricle & 2845 \\
Right Parietal Operculum Cortex & 2743 \\
Left Central Operculum Cortex & 2728 \\
Left Pallidum & 2705 \\
Right Parahippocampal Gyrus & 2684 \\
Left Thalamus & 2637 \\
\hline
\end{tabular}
\end{table}

Notably, the SHAP analysis ranks CBV as the least import image type, despite statistical tests in Tables~\ref{tab:boxplot_summary} and~\ref{tab:asymmetry_results} identifying it as second most important out of four image types. Similarly, PMDs related to unimodality were ranked highly by SHAP, even though they showed limited discriminative power in the univariate analyses of Tables~\ref{tab:boxplot_summary} and~\ref{tab:asymmetry_results}.

This discrepancy likely reflects methodological differences: logistic regression evaluates all features jointly, capturing interactions, while traditional statistical tests assess individual PMDs in isolation. This distinction is underscored by the SHAP value distribution—only 34\% of the total SHAP mass is concentrated in the top 500 features—suggesting that prediction depends on the aggregate influence of many weakly informative features. These findings highlight the potential of more advanced AI models to better capture complex patterns and leverage richer spatial information beyond the current PMD pipeline.

\subsection{Exploratory Analysis of Alternative Machine Learning Models}
\label{sec:other_ml_models}
In addition to the logistic regression model, we explored a range of alternative machine learning classifiers. We acknowledge that combining leave-one-out cross-validation with model selection constitutes a form of implicit data leakage, as model choice could then be influenced by performance on the test data. However, these supplementary analyses are included for completeness and to provide an exploratory comparison of model behavior.

All models were trained using the same preprocessing and evaluation as the logistic regression model in the main text. Inverse frequency class weighting was used to address class imbalance. Unless otherwise specified, default hyperparameters were used. 

The random forest classifier exhibited mode collapse, consistently predicting only the majority class despite class weighting. A similar, though slightly less severe, trend was observed for the K-nearest neighbors (KNN) model. In contrast, the XGBoost tree-based classifier did not suffer from this issue and maintained a better balance between sensitivity and specificity. The linear support vector classifier (SVC) demonstrated performance comparable to the logistic regression model but did not surpass it. Notably, the XGBoost linear model uniquely outperformed others in identifying true seizure cases, achieving a sensitivity of 88\% at the cost of a low specificity of 60\%. Overall, logistic regression remained the best overall performer across metrics.

\begin{figure}[!htbp]
\includegraphics[width=\textwidth]{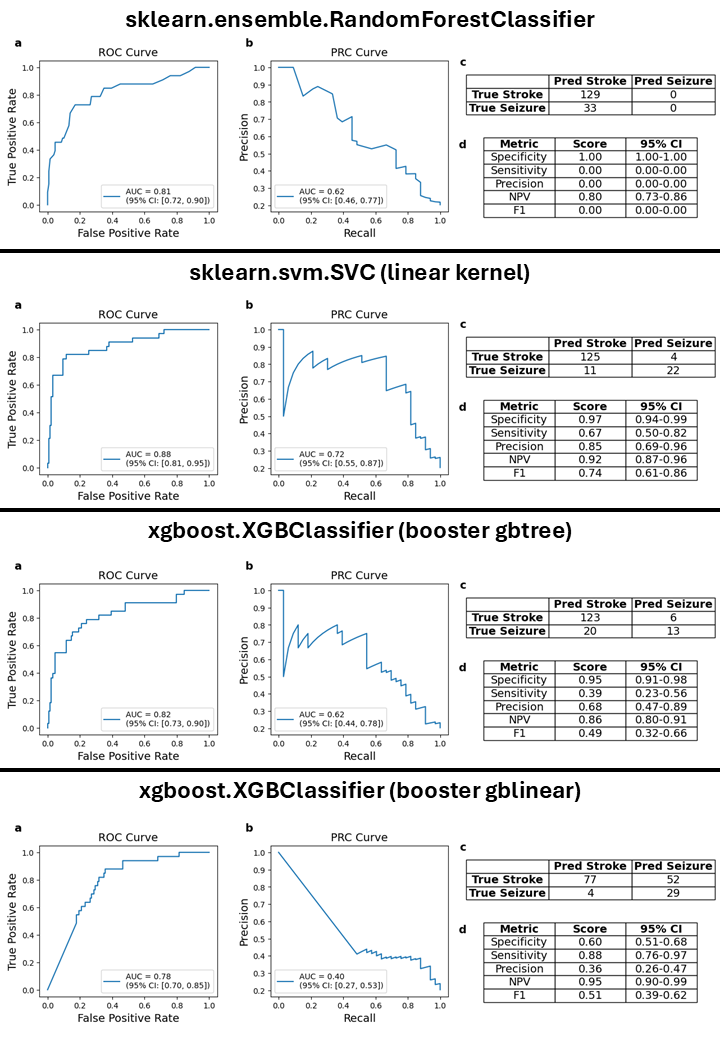}
\label{fig:ml_model_1}
\end{figure}

\begin{figure}[!htbp]
\includegraphics[width=\textwidth]{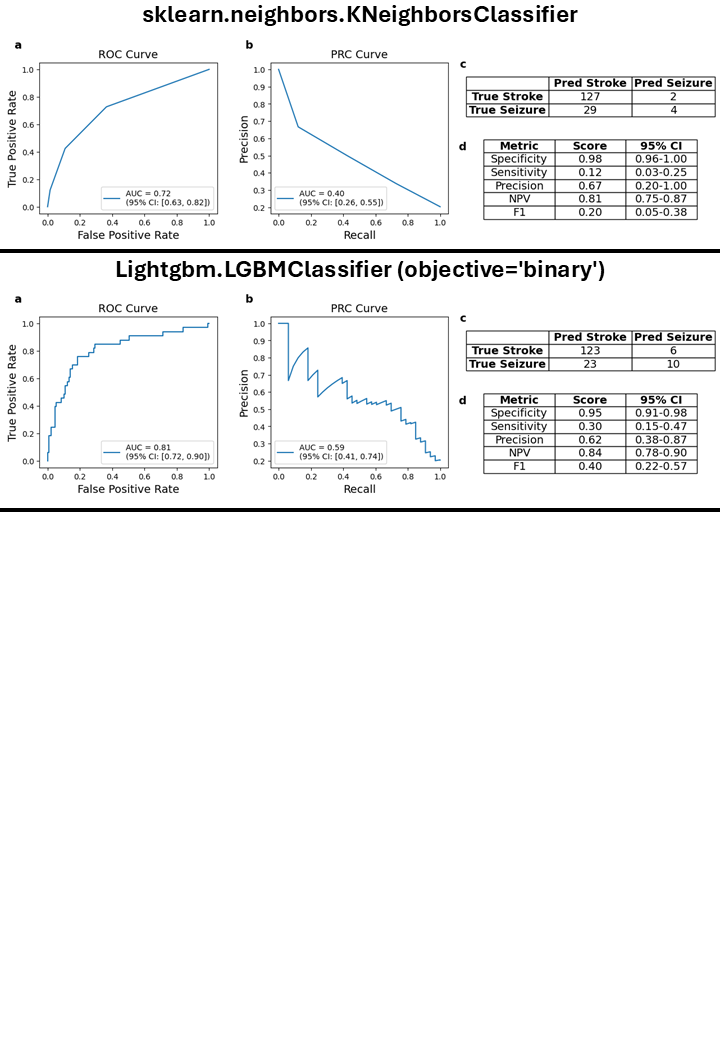}
\label{fig:ml_model_2}
\end{figure}